\documentclass[aps,prd,reprint,superscriptaddress,nofootinbib,amsmath,amssymb,floatfix]{revtex4-2}
\usepackage[T1]{fontenc}
\usepackage{graphicx}
\usepackage{bm}
\usepackage{booktabs}
\usepackage[colorlinks=true,linkcolor=blue,citecolor=blue,urlcolor=blue]{hyperref}
\graphicspath{{figures/}}

\newcommand{\dd}{\mathrm{d}}

\newcommand{\comp}{\mathrm{comp}}
\newcommand{\mref}{m_{\mathrm{ref}}}
\begin{document}
\raggedbottom
\title{Overlapping Subcritical Bubbles: Free Energy and Lifetime}

\author{Guangshang Chen}
\affiliation{
Key Laboratory of Particle Physics and Particle Irradiation (MOE),
Institute of Frontier and Interdisciplinary Science,
Shandong University, Qingdao 266237, P. R. China
}

\author{Yang Xiao}
\email{Corresponding author, xiaoyangphy@gmail.com}
\affiliation{
School of Physics, Henan Normal University,
Xinxiang 453007, P. R. China
}

\author{Jin Min Yang}
\affiliation{
School of Physics, Henan Normal University,
Xinxiang 453007, P. R. China
}
\affiliation{
Institute of Theoretical Physics,
Chinese Academy of Sciences,
Beijing 100190, P. R. China
}

\author{Yang Zhang}
\affiliation{
School of Physics, Henan Normal University,
Xinxiang 453007, P. R. China
}

\begin{abstract}
Subcritical bubbles can form an appreciable population during weak first-order phase transitions, but are usually treated as isolated fluctuations. 
This raises the question of whether spatial overlap between neighboring subcritical bubbles can modify their evolution. We address this question by combining analytic free-energy calculations for composite Gaussian profiles with Langevin simulations of overlapping configurations. 
We find that the overlap lowers the free-energy cost and generally increases the lifetime of subcritical bubbles, with the enhanced persistence potentially feeding back on their abundance.
Such overlap can therefore generate collective effects and should be incorporated into kinetic descriptions of subcritical-bubble populations.
\end{abstract}
\maketitle

\section{Introduction}
\label{sec:intro}
First-order phase transitions play an important role in the thermal history of the early Universe. At the electroweak scale, the observed Higgs mass of about \(125\,\mathrm{GeV}\) implies that electroweak symmetry breaking in the Standard Model (SM) proceeds through a smooth crossover rather than a first-order transition~\cite{Kajantie:1996mn,DOnofrio:2015gop,Laine:2015kra}. This situation can change in extensions of the SM. In particular, additional scalar degrees of freedom can modify the finite-temperature effective potential by enhancing cubic terms and thereby allow a first-order electroweak phase transition~\cite{Profumo:2007wc,Ramsey-Musolf:2019lsf,Quiros:1999jp}. Cosmological first-order phase transitions are therefore closely connected to physics beyond the SM and can have rich phenomenological consequences. Such transitions can provide the out-of-equilibrium condition required for electroweak baryogenesis~\cite{Morrissey:2012db,Cohen:1993nk} and can generate a stochastic gravitational wave background that may be observed by future space-based interferometers~\cite{Caprini:2015zlo,Caprini:2019egz,amaro2017laser,luo2016tianqin,ruan2020taiji}. Depending on the underlying model and transition dynamics, first-order phase transitions may also affect dark matter production~\cite{Li:2026qvn,Baker:2019ndr, Chway:2019kft,Xiao:2022oaq,Xiao:2023dbb}, primordial black hole formation~\cite{Liu:2021svg,Gouttenoire:2023naa,Ai:2026zrs}, and other relics of the early Universe~\cite{Hong:2020est,Krylov:2013qe,Jiang:2024zrb}. These connections have made cosmological first-order phase transitions an active subject at the interface of particle physics and early Universe cosmology.

The conventional description of a first-order phase transition is based on the nucleation of a critical bubble, which corresponds to a saddle-point configuration of the finite-temperature Euclidean action~\cite{Quiros:1999jp,Dine:1992wr,Coleman:1977py,Linde:1981zj}. A thermal fluctuation that reaches this critical configuration can subsequently grow, whereas smaller fluctuations tend to collapse back into the symmetric phase and are therefore referred to as subcritical bubbles~\cite{Gelmini:1992yj}. The possible role of such configurations was recognized early in studies of weak first-order transitions, where lowering the barrier between the two phases was found to increase the importance of these thermal fluctuations~\cite{Gleiser:1992ch}. As these fluctuations become easier to excite, they can occupy an appreciable fraction of the volume and drive the initially metastable background toward phase mixing~\cite{Gleiser:1994cf,Borrill:1994nk}. This picture was subsequently quantified through kinetic descriptions of the production, shrinkage, and size distribution of subcritical bubbles~\cite{Gelmini:1992yj}, and was later supported by stochastic field simulations, which found substantial phase mixing in weak transitions but a largely localized background in strong ones~\cite{Borrill:1994nk}. An analytic subcritical bubble model was further shown to reproduce the numerical phase fractions quantitatively and to relate complete phase mixing to the onset of percolation~\cite{Gleiser:1995er}. In the electroweak phase transition, related studies showed that thermal fluctuations can also limit the validity of the one-loop effective potential, while subcritical fluctuations may promote phase mixing and percolation~\cite{Gleiser:1992ed,Ramos:1996at}.

Previous studies therefore established that a substantial population of subcritical bubbles can arise in sufficiently weak first-order transitions. 
Our recent work further showed that a sizable subcritical bubble population can remain present near the nucleation temperature \(T_n\) even after cosmic expansion is taken into account~\cite{Chen:2026fap}. This makes the collective behavior of these fluctuations particularly relevant. Indeed, the induced nucleation scenario proposed that a critical bubble may be built up through the accumulation of shrinking subcritical bubbles, showing that neighboring fluctuations can play an important role in the transition~\cite{Enqvist:1994sj}.
However, although induced nucleation treated the accumulation of subcritical bubbles probabilistically and recognized the possibility of overlap, it did not address how such overlap modifies the underlying field configuration or its subsequent dynamics.
Once two subcritical bubbles overlap, the resulting configuration is no longer a simple collection of isolated fluctuations, and its energetic cost and persistence may consequently change. The present work therefore investigates how the overlap of subcritical bubbles modifies their free-energy cost and subsequent dynamics.

In this work, we investigate this problem through an analytic treatment of overlapping profiles and Langevin simulations, and find that overlap reduces the free-energy cost of the composite configurations and extends their persistence relative to the corresponding isolated subcritical bubbles. The remainder of the paper is organized as follows. Sec.~\ref{sec:model} presents the theoretical framework and analytic results. Sec.~\ref{sec:protocol} presents the numerical simulations and their results. Sec.~\ref{sec: conclusion} summarizes the main conclusions.

\section{Overlapping Configurations and Free Energy}
\label{sec:model}
In this work, we consider the benchmark single-field model with finite-temperature potential:
\begin{equation}\label{eq:potential}
    V(\phi,T) = D(T^2-T_0^2)\phi^2 - ET\phi^3 + \frac\lambda4\phi^4.
\end{equation}
The model parameters are phenomenologically determined and set to be $D=0.15$, $E=0.025$, $\lambda=0.10$ and $T_0=246\sqrt{\lambda/(2D)}$. At high temperatures, this potential features a unique minimum at $\phi_s = 0$, corresponding to the symmetric vacuum. As the temperature decreases, a secondary stationary point $\phi_b\neq 0$ develops below 
\begin{equation}
    T_1=T_0\sqrt{\frac{8\lambda D}{8\lambda D-9E^2}}.
\end{equation}
Upon further cooling, this state becomes the energetically favored true vacuum below the critical temperature
\begin{equation}
    T_c=\frac{T_0}{\sqrt{1-E^2/(\lambda D)}},
\end{equation}
thereby preparing the first-order phase transition. 

Within the interval $T_c<T<T_1$, subcritical bubbles emerge as localized fluctuations of the stable symmetric background toward the metastable broken-phase minimum. An isolated subcritical bubble centered at $\boldsymbol{x}_i$ can be modeled by configuration $\Phi_1(\boldsymbol{x}) = \phi_b g_i(\boldsymbol{x})$ with $g_i$ the Gaussian function~\cite{Gleiser:1995tv}
\begin{equation}
    g_i(\boldsymbol{x}) = \exp\left(-\frac{|\boldsymbol{x}-\boldsymbol{x}_i|^2}{R_i^2}\right).
\end{equation}
The width $R_i$ is not a sharp interface radius but the radius at which a constituent Gaussian falls to $e^{-1}$ of its central value. When $N$ bubbles coexist within the lifetime windows, we introduce the following ansatz via the inclusion-exclusion principle: 
\begin{equation}\label{eq:ansatz}
\begin{aligned}
    \frac{\Phi_N(\bm x)}{\phi_b} &= 1-\prod_{i=1}^N(1-g_i(\bm x)) \\
&= \sum_{\bm\sigma}(-)^{|\bm\sigma|+1}\prod_i g_i^{\sigma_i}(\bm x),
\end{aligned}
\end{equation}
where $\bm \sigma=(\sigma_1,\ldots,\sigma_N)$ runs over all the non-vanishing $N$-tuple of Boolean value $\sigma_i\in\{0,1\}$. 

Such a configuration provides a natural smooth-union prescription for overlapping localized profiles which also admits a simple probabilistic motivation. The single Gaussian function $g_i\in[0,1]$ is regarded as the local occupation weight associated with an independently sampled local bubble-forming event. The probability that point $\bm x$ remains unconverted is thus the product $\prod_i(1-g_i)$. Consequently, Eq.~\eqref{eq:ansatz} gives the complementary occupation weight, preserves $0\leq\Phi_N\leq\phi_b$, and avoids producing an arbitrarily large central amplitude between the overlap constituents. This probabilistic picture is used only to motivate the composition rule and does not identify the classical field profile with a microscopic tunneling probability. It is a prescribed family of profiles, but not a family of stationary solutions.

Correspondingly, the static free energy associated with the profile in Eq.~\eqref{eq:ansatz} is
\begin{equation}\label{eq:functional}
\begin{aligned}
    F_N &= \int_{\mathbb R^3}\dd^3x \left[\frac12|\bm\nabla\Phi_N|^2+V(\Phi_N,T)\right].
\end{aligned}
\end{equation}
Both contributions can be evaluated analytically in terms of Gaussian overlap integrals:
\begin{equation}
    I_{\bm m}\equiv\int\dd^3x\prod_i g_i^{m_i}
    = \left(\frac{\pi}{A_{\bm m}}\right)^{3/2}e^{-Q_{\bm m}}.
\end{equation}
Here the $N$-tuple $\bm m\in \mathbb N^N$. In what follows, we reserve the Greek indices for Boolean values and use the Latin indices for integer values. The parameters $A_{\bm m}$ and $Q_{\bm m}$ are determined via the weights $w_i=m_i/R_i^2$ that
\begin{align}
    A_{\bm m} &= \sum_iw_i, &
    Q_{\bm m} &= \frac{\sum_{i<j}w_iw_j d_{ij}^2}{A_{\bm m}}.
\end{align}
The separation matrix $d_{ij}=|\bm x_i-\bm x_j|$ denotes the separation between the centers of the \(i\)-th and \(j\)-th bubbles. It provides another geometrical characterization of the multiple-bubble configuration. 

For the kinetic term, differentiation of the Gaussian profiles generates only polynomial prefactors multiplying products of Gaussians, and the resulting integrals are therefore standard Gaussian moments. For a polynomial effective potential, the potential term likewise reduces to products of Gaussian profiles. Concretely, we arrive at 
\begin{eqnarray}
    F_N &=& \phi_b^2\left(\sum_{\bm\sigma_1,\bm\sigma_2}(-)^{|\bm\sigma_1|+|\bm\sigma_2|}\mathcal H_{\bm\sigma_1\bm\sigma_2}+ D(T^2-T_0^2)\mathcal M_2\right) \nonumber \\
    && \quad\ - ET \mathcal M_3\phi_b^3 +  \frac{\lambda}{4}\mathcal M_4\phi_b^4.
\end{eqnarray}
Here, the several moments are defined to be
\small
\begin{eqnarray}
    && \mathcal H_{\bm\sigma_1\bm\sigma_2} = \frac12\int\dd^3x\, \nabla\left(\prod_i g_i^{\sigma_{1i}}\right)\cdot\nabla\left(\prod_j g_j^{\sigma_{2j}}\right) \nonumber \\
    &&= I_{\bm\sigma_1+\bm\sigma_2}\left[\frac{3A_{\bm\sigma_1}A_{\bm\sigma_2}}{A_{\bm\sigma_1}+A_{\bm\sigma_2}}-\frac{2A^2_{\bm\sigma_1}A^2_{\bm\sigma_2}|\bm\mu_{\bm\sigma_1}-\bm\mu_{\bm\sigma_2}|^2}{(A_{\bm\sigma_1}+A_{\bm\sigma_2})^2}\right]
\end{eqnarray} 
\normalsize
where the weighted center $\bm\mu_{\bm\sigma} = \sum_{i=1}^Nw_i\bm x_i/A_{\bm\sigma}$ and 
\begin{equation}
    \mathcal M_q = \sum_{\bm\sigma_1...\bm\sigma_q}(-)^{|\bm\sigma_1| +...+|\bm\sigma_q|+q}I_{\bm\sigma_1 +...+ \bm\sigma_q}.
\end{equation}

If one isolates the $i$-th bubble from the cluster, the free energy of the single Gaussian configuration is 
\begin{equation}\label{eq:single}
\begin{aligned}
    F_1(R_i) = \frac{\pi^{3/2}R_i^3}{2^{3/2}}\bigg[&\frac{3\phi_b^2}{2R_i^2} + D(T^2-T_0^2)\phi_b^2 \\
    &\ - \frac{2^{3/2}}{3^{3/2}}ET\phi_b^3 + \frac{\lambda\phi_b^4}{2^{7/2}}\bigg].
\end{aligned}
\end{equation}
We quantify the effect of bubble overlap by defining the excess free energy
\begin{equation}
\Delta F_N=F_N-\sum_{i=1}^NF_i.
\label{eq:deltaF}
\end{equation}
This quantity measures the total free-energy correction induced by bubble overlap relative to isolated constituents. It should be distinguished from an irreducible $N$-body interaction, as $\Delta F_N$ contains all overlap contributions up to order $N$.

For simplicity, we restrict our analysis to symmetric configurations characterized by a common nearest-neighbor separation distance $d$. Depending on the geometry, $d$ corresponds to the separation of a bubble pair, the spacing along collinear three bubbles, or the edge length of an equilateral triangle, square, or regular tetrahedron. Illustrations of these configurations can be found in the figure below. When $d$ becomes large compared with the bubble width, all mixed Gaussian integrals vanish and $\Delta F_N\to0$. The same conclusion could also be drawn by reviewing the configuration ansatz in Eq.~\eqref{eq:ansatz}. For large separation distance, 
\begin{equation}
    \frac{\Phi_N(\bm x)}{\phi_b} \approx \sum_{i=1}^Ng_i(\bm x),
\end{equation}
which preserves the additivity of the free energy as expected for a collection of independent bubbles.

\begin{figure*}[t]
\includegraphics[width=\textwidth]{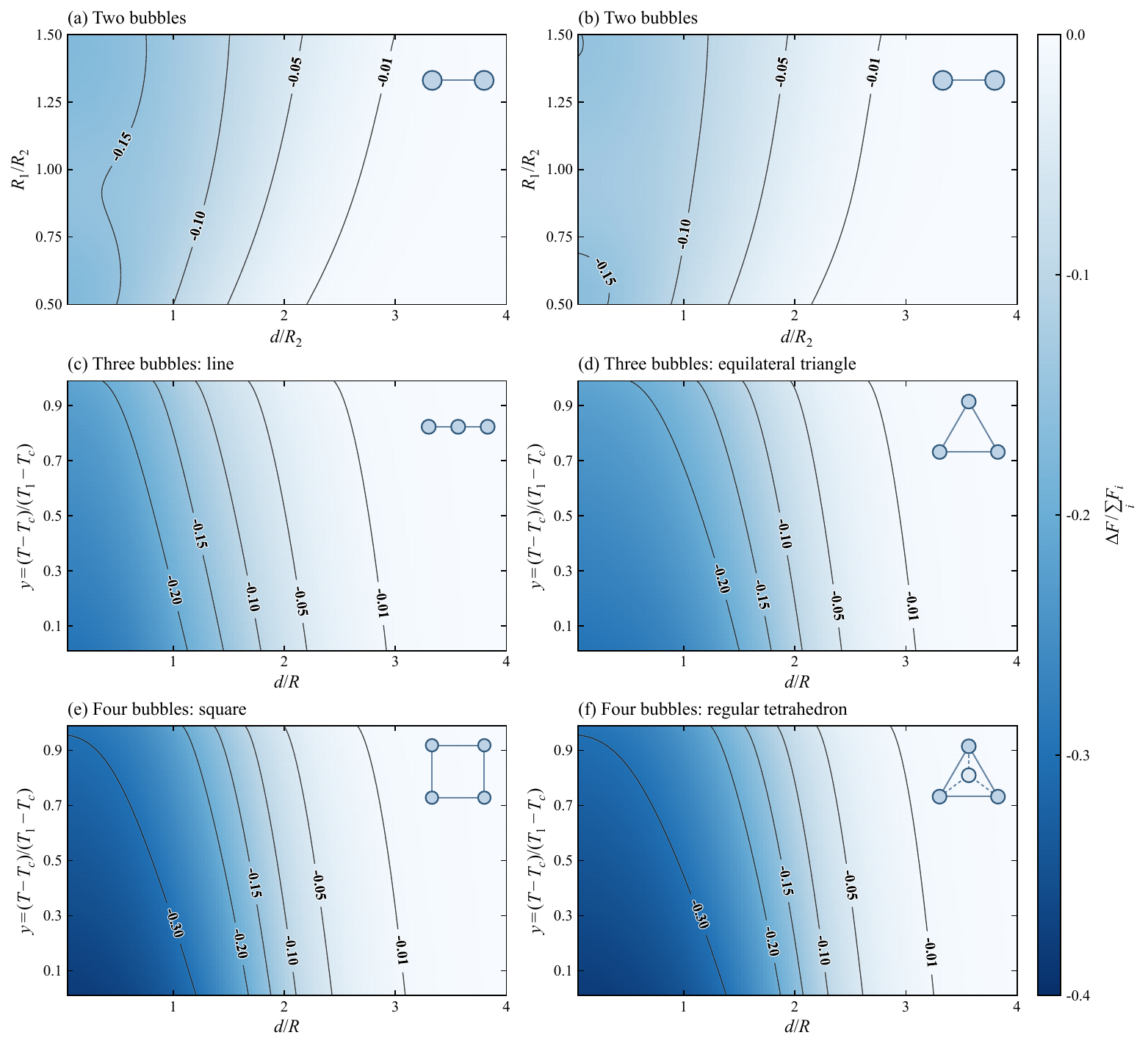}
\caption{Normalized free-energy difference
$\Delta F_N/\sum_i F_i$ between composite Gaussian configurations
and their isolated constituents.
Panels (a) and (b) show the two-bubble scans at $y=1/3$ and $2/3$,
respectively, with fixed $R_2=R$.
Panels (c)--(f) show equal-width multibubble configurations over
$0.01\leq y\leq0.99$, where $y=(T-T_c)/(T_1-T_c)$.
The separation $d$ denotes the adjacent spacing of the line
or the edge length of the other multibubble geometries.
Colors and labeled white contours indicate the same normalized
free-energy difference.}
\label{fig:static}
\end{figure*}


For a comprehensive view, Fig.~\ref{fig:static} illustrates the normalized excess free energy 
\begin{equation}
    \epsilon_F=\frac{\Delta F_N}{\sum_iF_i}
\end{equation}
for both unequal-width bubble pairs and multi-bubble geometries. Across all examined configurations, the interaction energy is consistently negative and asymptotically vanishes as the spatial separation increases. For the two-bubble case, the free-energy deficit reaches approximately $17\%$ and exhibits a strong sensitivity to the radius ratio $R_1/R_2$, which modulates both the isolated constituent energies and their spatial overlap profile. Extending the calculation to symmetric multi-bubble arrangements over $0.01 \leq y \leq 0.99$, the maximum energy reductions deepen to around $30\%$ for three constituents and $38\%$ for four. At a fixed nearest-neighbor spacing, geometries with higher coordination numbers exhibit systematically more pronounced reductions at moderate separations—the triangular configuration shows a larger deficit than the collinear chain, while the regular tetrahedron exceeds the square. 

A negative \(\Delta F_N\) indicates that the prescribed overlapping configuration has a lower free energy than the corresponding set of isolated Gaussian bubbles with identical individual parameters. Thus, \(\Delta F_N\) quantifies the energetic modification induced by finite spatial overlap and shows that such composite configurations can be energetically favored relative to the dilute-bubble picture. This energetic comparison, however, only establishes the possibility of these configurations and does not determine whether they can persist under the full field dynamics. This is intrinsically a dynamical question, which we investigate in the next section through explicit real-time evolution.

\section{Numerical Simulation of the Langevin Equation}
\label{sec:protocol}
\subsection{Langevin dynamics and numerical setup}

Subcritical bubbles are finite-amplitude thermal fluctuations of the
order-parameter field. Their real-time evolution in a thermal environment
can therefore be described by a coarse-grained Langevin equation,
\begin{align}
    \frac{\partial \phi}{\partial t} &= \pi ,
    \notag \\
    \frac{\partial \pi}{\partial t}
    &=
    \nabla^2 \phi
    - \frac{\partial V(\phi,T)}{\partial \phi}
    - \eta \pi
    + \xi(\mathbf{x},t),
    \label{eq:langevin_dimensional}
\end{align}
where $\eta$ is the damping coefficient and $\xi$ represents the thermal
noise~\cite{Borrill:1994nk}. We model the thermal noise as Gaussian white noise satisfying the fluctuation-dissipation relation
\begin{equation}
    \left\langle
    \xi(\mathbf{x},t)\,
    \xi(\mathbf{x}',t')
    \right\rangle
    =
    2\eta T\,
    \delta^{(3)}(\mathbf{x}-\mathbf{x}')
    \delta(t-t').
    \label{eq:noise_dimensional}
\end{equation}
The stochastic dynamics defined above should be understood as a coarse-grained description of the long-wavelength thermal field, rather than as a microscopic theory valid at arbitrarily short distances~\cite{Greiner:1996dx,Gleiser:1993ea}. This effective description therefore requires a finite ultraviolet cutoff. In the numerical evolution, we retain only Fourier modes with $|\mathbf{k}| \leq \Lambda$, so that fluctuations below the corresponding length scale are excluded from both the stochastic sampling and the subsequent evolution. This mode restriction is also imposed consistently on the scalar field and all terms entering its stochastic evolution, ensuring that the dynamics remain within the same coarse-grained description.

For the numerical calculation, we normalize all dimensional quantities by a
fixed reference mass scale,
\begin{equation}
    T_{\rm ref}=\frac{T_c+T_1}{2},
    \qquad
    m_{\rm ref}^2
    \equiv
    \left.
    \frac{\partial^2 V(\phi,T)}{\partial\phi^2}
    \right|_{\phi=0,\,T=T_{\rm ref}}. 
    \label{eq:mref}
\end{equation}

For any quantity $Q$ with mass dimension $d_Q$, we define
$\tilde Q=Q/m_{\rm ref}^{d_Q}$. For example,
\begin{equation}
    \tilde{\mathbf{x}}=m_{\rm ref}\mathbf{x},
    \quad
    \tilde t=m_{\rm ref}t,
    \quad
    \tilde\phi=\frac{\phi}{m_{\rm ref}},
    \quad
    \tilde T=\frac{T}{m_{\rm ref}} .
    \label{eq:dimensionless_variables}
\end{equation}
Unless otherwise stated, the simulations use
\begin{equation}
\begin{aligned}
    \tilde R &= 3,\qquad
    \tilde L = 24,\qquad
    \tilde\eta = 1, \\
    \tilde\Lambda &= 2,\qquad
    \Delta\tilde x = 0.375,\qquad
    \Delta\tilde t = 0.0125 .
\end{aligned}
\label{eq:numerical_parameters}
\end{equation}
on a periodic $64^3$ lattice. The spatial derivatives are evaluated using
the nearest-neighbor lattice Laplacian, while the time evolution is performed
with the BAOAB splitting scheme~\cite{leimkuhler2013rational,leimkuhler2013robust}. The temperature is specified by the
dimensionless variable $y = (T-T_c)/(T_1-T_c)$. Details of the numerical
implementation and the stability and robust tests are given in
Appendix~\ref{app:numerics}.

\subsection{Numerical framework for configuration lifetime analysis}

Having specified the stochastic dynamics and numerical setup, we now turn to
the lifetime analysis of the composite configurations. We investigate how the
persistence depends on the separation $d$, the constituent widths $R_i$, and
the spatial arrangement of the bubbles. In addition to two-bubble
configurations, we consider three bubbles arranged in a line or an
equilateral triangle and four bubbles arranged in a square or a regular
tetrahedron. Temperature is varied independently to probe how these geometric
effects change across the metastable regime.

For each parameter choice, we perform 30 independent experiments with
different random seeds. In each experiment, the composite configuration and
its $N$ isolated constituents are evolved separately under the same
realization of the thermal noise. Each isolated branch retains the location, the width and
amplitude of the corresponding constituent in the composite configuration,
so that the comparison isolates the effect of combining the constituents
into a common field configuration. 

We define the lifetime by tracking the persistence of the field region that
initially lies above the barrier,
$\phi(\mathbf{x},t)>\phi_{\rm top}(T)$. During the evolution, connected
above-barrier regions associated with the initial configuration are followed,
while brief stochastic interruptions are not immediately identified as
collapse. For an isolated constituent, this procedure tracks the persistence
of its initial above-barrier region, whereas for a composite configuration
all initially tagged regions are treated as a single tracked configuration.
The same tracking prescription is therefore applied to both the isolated and
composite branches, yielding their respective lifetimes. Further details of
the tracking procedure are given in Appendix~\ref{app:numerics}.

\subsection{Lifetime results}
\begin{figure*}[t]
\includegraphics[width=\textwidth]{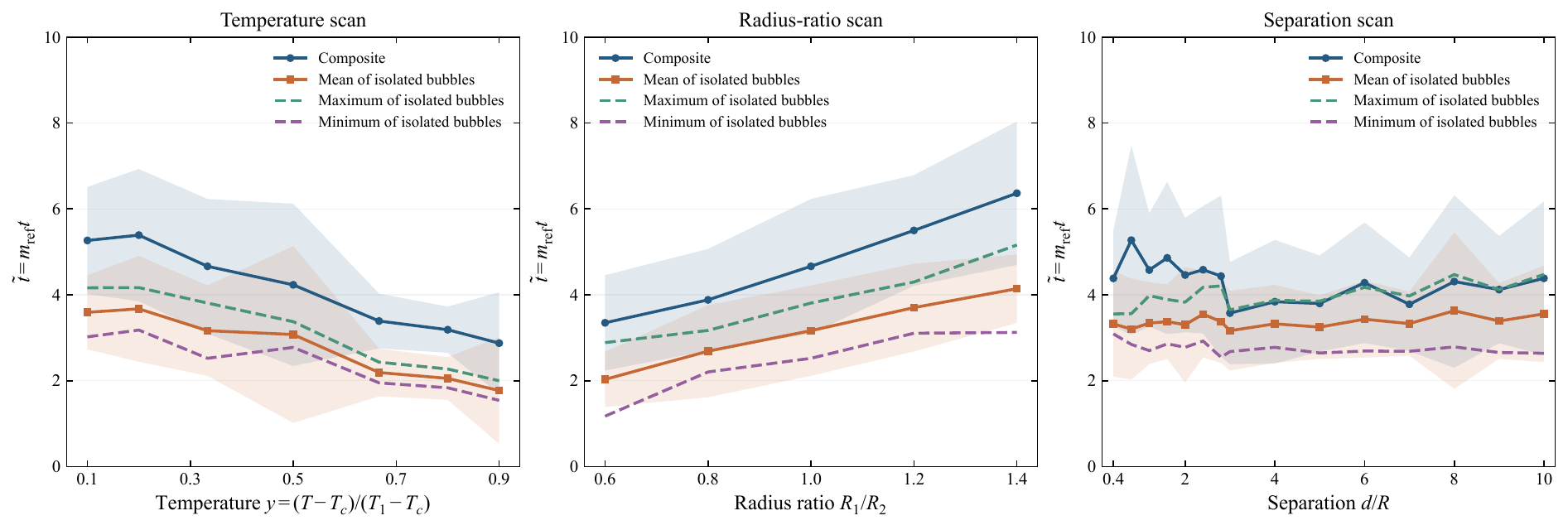}
\caption{
Two-bubble lifetimes, expressed as $\tilde{t}=m_{\mathrm{ref}}t$,
as functions of temperature (left), radius ratio (middle),
and separation (right).
The temperature scan uses equal widths and $d/R=0.8$.
The radius-ratio scan fixes $y=1/3$, $R_2=R$, and $d/R_2=0.8$,
while the separation scan uses equal widths at $y=1/3$.
Solid curves compare the composite lifetime with the
within-experiment mean isolated lifetime.
Dashed curves show the ensemble averages of the within-experiment
maximum and minimum isolated lifetimes.
Each point averages 30 stochastic experiments and shaded bands span one sample standard deviation about the solid curves.
}
\label{fig:pairlife}
\end{figure*}

Fig.~\ref{fig:pairlife} compares the lifetime of the two-bubble
configuration with those of its isolated constituents as functions of
temperature, radius ratio, and separation. The solid blue curve shows the
ensemble-averaged lifetime of the composite configuration, while the
corresponding shaded band denotes one sample standard deviation over 30
independent stochastic realizations.

To construct the isolated reference, each constituent is evolved separately
while retaining the position it occupies in the composite configuration.
Although all branches within a given experiment are driven by the same
spatial noise realization, the local fluctuations sampled at the two bubble
locations are generally different. Consequently, even two isolated bubbles
with equal radii need not have identical lifetimes in a single realization.
For noise realization $r$, we denote the lifetime of the $i$-th isolated constituent by $\tilde t_{i,r}$ and that of the composite configuration by
$\tilde t_{\comp,r}$. To characterize the variation of isolated bubbles, we define
\begin{equation}
    \begin{aligned}
    \tilde t_{\max,r}
    &=
    \max(\tilde t_{1,r},\tilde t_{2,r}),
    \\
    \tilde t_{\min,r}
    &=
    \min(\tilde t_{1,r},\tilde t_{2,r}) .
\end{aligned}
\end{equation}
The green and purple dashed curves show the ensemble averages of these maximum and minimum lifetimes, respectively, whereas the
orange curve gives the ensemble-averaged mean isolated lifetime. Since the longer-lived isolated constituent provides a more stringent reference for assessing the effect of overlap on the lifetime, we use the maximum isolated lifetime as the baseline in the subsequent analysis.

We first consider the temperature dependence shown in the left panel. The
lifetimes generally increase as the temperature approaches $T_c$ from above.
In this regime, the interior of a subcritical bubble lies near the metastable
nonzero minimum, whose free-energy density approaches that of the surrounding
symmetric phase as $T\to T_c^+$. Accordingly,
\begin{equation}
    \Delta V(T)
    =
    V(\phi_b,T)-V(0,T)
    >0
\end{equation}
decreases toward coexistence, thereby weakening the inward bulk driving
associated with
$p_{\mathrm{in}}-p_{\mathrm{out}}=-\Delta V$.
This provides a simple thermodynamic interpretation of the enhanced
persistence near $T_c$, although the potential barrier and the strength of
thermal fluctuations also vary simultaneously with temperature. Importantly,
throughout the sampled temperature range, the composite configuration remains
longer lived on average than both the mean and the maximum isolated
references.

The middle panel isolates the effect of the constituent size by varying
$R_1/R_2$ at fixed $R_2$. As $R_1$ increases, both the composite and isolated
reference lifetimes increase. For the isolated subcritical bubble, this trend is mainly driven by the longer persistence of the enlarged constituent. As the bubble radius increases, the effect of surface tension $\sigma$ becomes weaker, so that the tendency of the bubble to shrink is reduced. In the thin-wall approximation, this contribution to the inward pressure scales as \(2\sigma/R\). Although the Gaussian profiles considered here do not possess a sharp interface, this relation still provides a useful qualitative interpretation of the observed radius dependence. The composite lifetime shows a similar, approximately linear increase with \(R_1/R_2\) over the displayed range and remains systematically above the isolated references.

The separation scan in the right panel provides a more direct evidence of
the role of overlap. Under the adopted tracking prescription, a composite
configuration is regarded as surviving as long as at least one inherited
above-threshold region remains. Therefore, when the bubbles are sufficiently far
apart and their local evolution becomes effectively independent, the
composite lifetime is expected to approach that of the last surviving
isolated constituent
\begin{equation}
    \lim_{d \to \infty}\left[\tilde t_{\mathrm{comp},r}-\tilde t_{\max,r}\right] = 0.
\end{equation}
This behavior is indeed observed at large separation. For example, at $d/R=10$, their
mean lifetimes are approximately $4.38$ and $4.47$, respectively. The close
agreement provides a useful consistency check of the separated-bubble limit
under the adopted lifetime definition. 

The behavior changes qualitatively once the bubbles begin to overlap. At
small separation $d < 2R$, the mean lifetime of the composite configuration exceeds
even that of the longer-lived isolated constituent. This excess cannot be
explained solely by the statistical advantage of selecting the last survivor
from two independently evolving bubbles. Instead, it indicates that the
overlap-modified field configuration and its subsequent coupled evolution
provide an additional enhancement of the persistence. 

\begin{figure*}[t]
\includegraphics[width=0.87\textwidth]{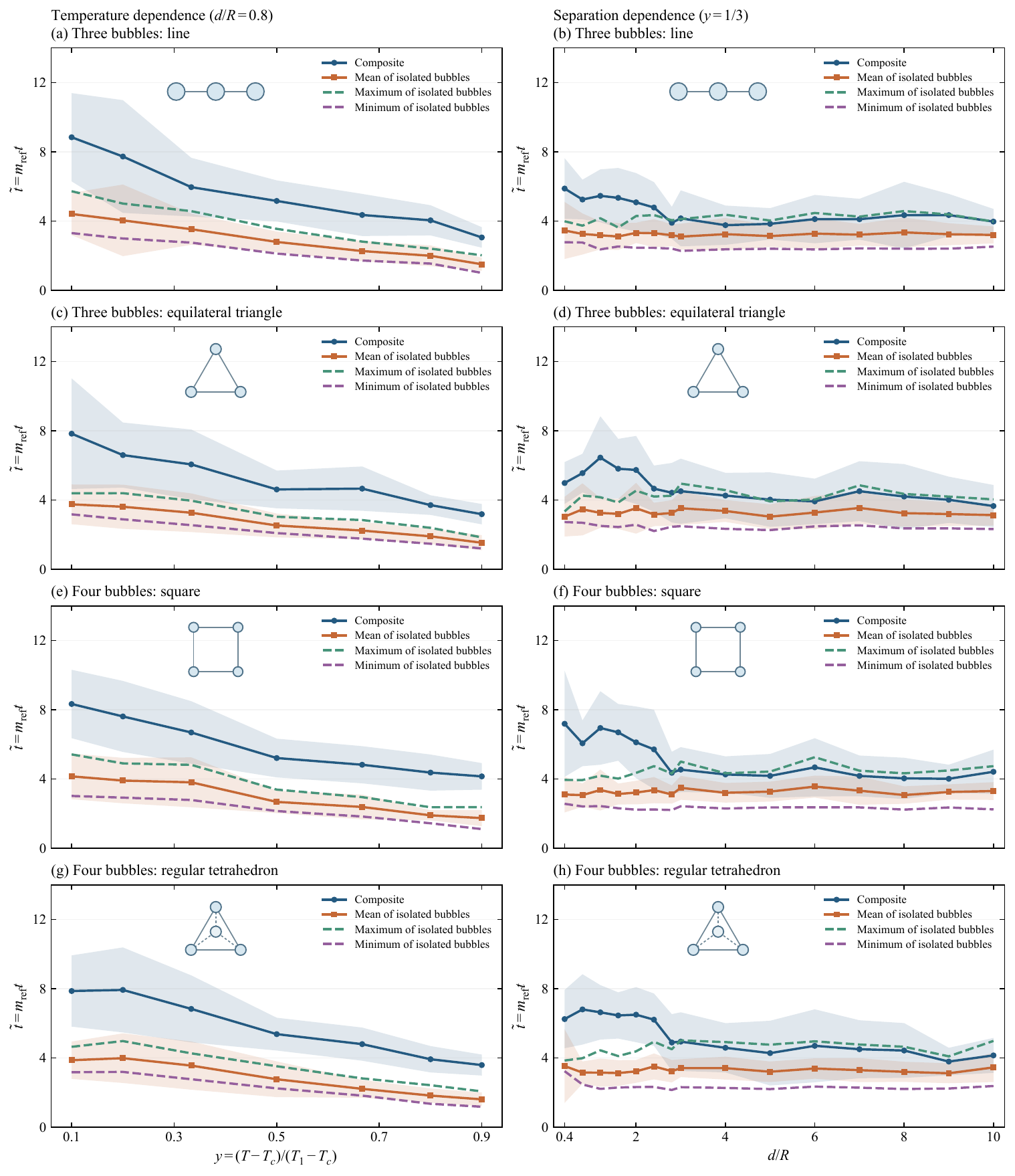}
\caption{
Mean dimensionless lifetimes of equal-width multibubble configurations,
with each row corresponding to the geometry illustrated in the insets.
The left column varies temperature at $d/R=0.8$,
and the right column varies separation at $y=1/3$.
On the right panel, the maximum isolated lifetime may exceed the composite lifetime at large separation. This residual mismatch arises from limited statistics and the bubble tracking procedure, and can be reduced by increasing the sample size and refining the tracking window.}
\label{fig:multitemp}
\end{figure*}

The same qualitative comparison extends beyond two constituents. Fig.~\ref{fig:multitemp} shows that the line, triangle, square, and tetrahedron all retain a positive difference between the composite and the isolated lifetime at each temperature and small separation. 

We next examine how the spatial configuration affects the composite lifetime. Comparing Fig.~\ref{fig:pairlife} and Fig.~\ref{fig:multitemp} shows that multibubble configurations generally persist longer than the pair in the regime of substantial overlap. In the temperature scans at \(d/R=0.8\), the composite mean lifetimes of the three-bubble line and triangle exceed those of the pair by approximately \(6\%\)–\(68\%\) and \(9\%\)–\(49\%\), respectively. The corresponding increase for the four-bubble configurations is approximately \(23\%\)–\(58\%\). These comparisons indicate that involving additional overlapping constituents can enhance persistence, although the increase is not strictly ordered by \(N\). At the lowest sampled temperature, for example, the three-bubble line has a larger mean lifetime than either four-bubble configuration.

The two three-bubble arrangements also exhibit distinct behavior in the overlap regime. In the separation scan at \(y=1/3\), the mean lifetime of the triangle exceeds that of the line by approximately \(6\%\)–\(18\%\) over the sampled range \(0.8\leq d/R\leq2\), whereas the ordering reverses at the smallest separation, \(d/R=0.4\), where the line becomes longer lived. Both three-bubble configurations generally persist longer than the pair at these short separations, although the line and pair have nearly identical mean lifetimes at \(d/R=0.8\). Geometrically, the triangle contains three constituent pairs separated by \(d\), while the line contains only two such pairs, with the remaining pair separated by \(2d\). The reversal at the smallest separation therefore shows that a larger number of closely overlapping pairs does not guarantee a longer lifetime. 

A similar conclusion emerges from comparing the square and tetrahedral configurations at fixed \(N=4\). Their mean lifetimes remain comparable in the strongly overlapping regime, and their ordering changes even among the shortest separations: the square is longer lived at \(d/R=0.4\), whereas the tetrahedron becomes longer lived at \(d/R=0.8\). Thus, arranging the same number of constituents in a genuinely three-dimensional geometry does not provide a systematic persistence advantage over a planar configuration. 

Taken together, these results support enhanced persistence of overlapping multibubble configurations relative to the two-bubble case, but do not reveal a simple lifetime hierarchy determined solely by constituent number, geometric compactness, or the number of closely separated pairs.

The enhanced persistence found above raises a natural question: can the
overlap of nearby subcritical bubbles modify their abundance? As a reference,
we start from the one-body fast estimate introduced in
Ref.~\cite{Chen:2026fap}. In the full kinetic description, the
subcritical population is characterized by the radius distribution
$n(R,T)$, with the total number density
\begin{equation}
    \mu(T)=\int dR\,n(R,T).
\end{equation}
To obtain a simple one-body estimate, the radius dependence of the removal
process is represented by the characteristic scale
$\xi_s=[V''(0,T)]^{-1/2}$, corresponding to the symmetric-phase correlation
length. The quasistationary solution then gives
\begin{equation}
    \mu_{\rm qs}
    \simeq
    \frac{S_*}{K_\xi},
    \qquad
    V_\xi=\frac{4\pi}{3}\xi_s^3,
    \qquad
    \hat f_\xi
    =
    V_\xi\mu_{\rm qs}.
    \label{eq:dilute_reference}
\end{equation}
Here $\mu_{\rm qs}$ is the quasistationary total number density of
subcritical bubbles, $S_*$ is the effective production source in the fast
estimate, and $K_\xi$ is the one-body removal rate evaluated at the
representative scale $\xi_s$. The quantity $\hat f_\xi$ is the corresponding
additive volume estimate obtained by assigning the characteristic volume
$V_\xi$ to each bubble. Equivalently, introducing the one-body residence time
$\tau_1=K_\xi^{-1}$ gives
$\mu_{\rm qs}\simeq S_*\tau_1$, making explicit that the abundance depends
on both the production rate and the persistence of the fluctuations.

This one-body description assumes that neighboring bubbles do not modify one
another's evolution. To estimate when such effects may become relevant, we
introduce an interaction range
\begin{equation}
    d_{\rm int}=\kappa\xi_s .
\end{equation}
For a homogeneous Poisson population with number density $\mu_{\rm qs}$,
the mean number of neighboring bubbles within this range is simply the number
density multiplied by the interaction volume,
\begin{equation}
    \chi
    =
    \mu_{\rm qs}\frac{4\pi}{3}d_{\rm int}^3
    =
    \kappa^3\hat f_\xi,
    \qquad
    P_{\geq1}=1-e^{-\chi},
    \label{eq:pair_probability}
\end{equation}
where $P_{\geq1}$ is the probability that a given bubble has at least one
neighbor within $d_{\rm int}$. In the dilute limit,
$P_{\geq1}=\chi+O(\chi^2)$, so the fraction of bubbles entering a pair
environment is approximately
$q\simeq\kappa^3\hat f_\xi$. Notice that even when
$\hat f_\xi\ll1$, the pair probability can be appreciable if the interaction
range extends beyond the characteristic bubble radius.

To estimate how the persistence enhancement feeds back on the population,
we consider a minimal extension in which the production rate is unchanged,
while bubbles in a pair environment have a mean residence time
\begin{equation}
    \tau_{\rm env}
    =
    (1+\epsilon_{\rm res})\tau_1 ,
\end{equation}
where $\epsilon_{\rm res}$ parametrizes the relative enhancement of the
residence time. The corresponding quasistationary number density is
\begin{equation}
    \mu
    =S_*\left[(1-q)\tau_1+q\tau_{\rm env}\right]
    =\mu_{\rm qs}
    \left(
        1+\epsilon_{\rm res}q
    \right).
    \label{eq:birth_residence}
\end{equation}
Using $q=\kappa^3\hat f_\xi+O(\hat f_\xi^2)$, the additive volume estimate
becomes
\begin{equation}
    \hat f_\xi^{\rm corr}
    =
    \hat f_\xi
    +
    \epsilon_{\rm res}\kappa^3\hat f_\xi^2
    +
    O(\hat f_\xi^3).
    \label{eq:leading_population_correction}
\end{equation}
The leading correction is therefore quadratic in the dilute population, as
expected for an effect that requires two nearby fluctuations, while the
relative correction is
\begin{equation}
    \frac{
        \hat f_\xi^{\rm corr}-\hat f_\xi
    }{
        \hat f_\xi
    }
    =
    \epsilon_{\rm res}\kappa^3\hat f_\xi
    +
    O(\hat f_\xi^2).
\end{equation}
As an illustration, taking the approximately $48\%$ lifetime enhancement
found for the reference two-bubble configuration as a representative value
of $\epsilon_{\rm res}$, together with
$\hat f_{\xi,*}=10^{-1.95}$ and $\kappa=2$, gives a leading relative
correction of about $4.3\%$. This estimate is based on a dilute,
uncorrelated Poisson population. If the actual subcritical-bubble density
becomes large, higher-order clusters and spatial correlations can no longer
be neglected, and the simple quadratic correction in
Eq.~\eqref{eq:leading_population_correction} need not remain valid.

\section{conclusions}
\label{sec: conclusion}
In this work, we investigated the collective behavior of overlapping subcritical bubbles using both analytic free-energy calculations and Langevin simulations. We find that (1) overlap lowers the free-energy cost of the composite configuration and, at the same time, increases its persistence relative to isolated constituents; (2) A simple dilute estimate further shows that this enhanced persistence can feed back on the subcritical-bubble abundance, providing a leading correction to the conventional one-body description. 

These results suggest that overlap should be regarded as an intrinsic part of subcritical bubble dynamics rather than a purely geometric effect. A natural next step is therefore to develop a self-consistent kinetic framework that incorporates spatial correlations, overlap-dependent lifetimes, and higher-order clusters, allowing the abundance and collective evolution of subcritical bubbles to be treated on the same footing.

\appendix

\section{Numerical implementation and parameter robustness}
\label{app:numerics}

We use the dimensionless variables defined in
Eq.~\eqref{eq:dimensionless_variables}, with all evolution times and
lifetimes expressed as $\tilde t=\mref t$.
The field is discretized on a periodic lattice with the nearest-neighbor
Laplacian,
\begin{equation}
 (\widetilde\nabla_h^2\tilde\phi)_{\bm n}
 =\frac{1}{(\Delta\tilde x)^2}
 \sum_{j=1}^{3}
 \left(\tilde\phi_{\bm n+\bm e_j}
       +\tilde\phi_{\bm n-\bm e_j}-2\tilde\phi_{\bm n}\right).
 \label{eq:lattice_laplacian}
\end{equation}
Time integration uses the BAOAB scheme~\cite{leimkuhler2013rational}.
Its thermal substep updates $\tilde\pi=\pi/\mref^2$ according to
\begin{equation}
 \tilde\pi\longleftarrow
 e^{-\tilde\eta\Delta\tilde t}\tilde\pi+
 \sqrt{\frac{\tilde T(1-e^{-2\tilde\eta\Delta\tilde t})}
 {(\Delta\tilde x)^3}}\,P_{\tilde\Lambda}Z,
 \label{eq:OU}
\end{equation}
where $Z$ contains independent unit Gaussian variables on the lattice.
The projector $P_{\tilde\Lambda}$ retains modes with
$|\tilde{\bm k}|\leq\tilde\Lambda$ and is applied to the initial field,
nonlinear force, noise, and evolved field and momentum.
The temperature and width scans use the $64^3$ lattice specified in
Eq.~\eqref{eq:numerical_parameters}. The separation scans extending to
$d/R=10$ use a $320\times192\times192$ lattice with
$(\tilde L_x,\tilde L_y,\tilde L_z)=(120,72,72)$ at the same
$\Delta\tilde x=0.375$ and $\Delta\tilde t=0.0125$.

Lifetimes are measured every $\Delta\tilde t_{\rm obs}=0.025$ by following
connected above-barrier regions inherited from the initial configuration,
including connections across periodic boundaries. A nearby region may be
reacquired during a temporary absence. 

\begin{figure*}[t]
\includegraphics[width=\textwidth]{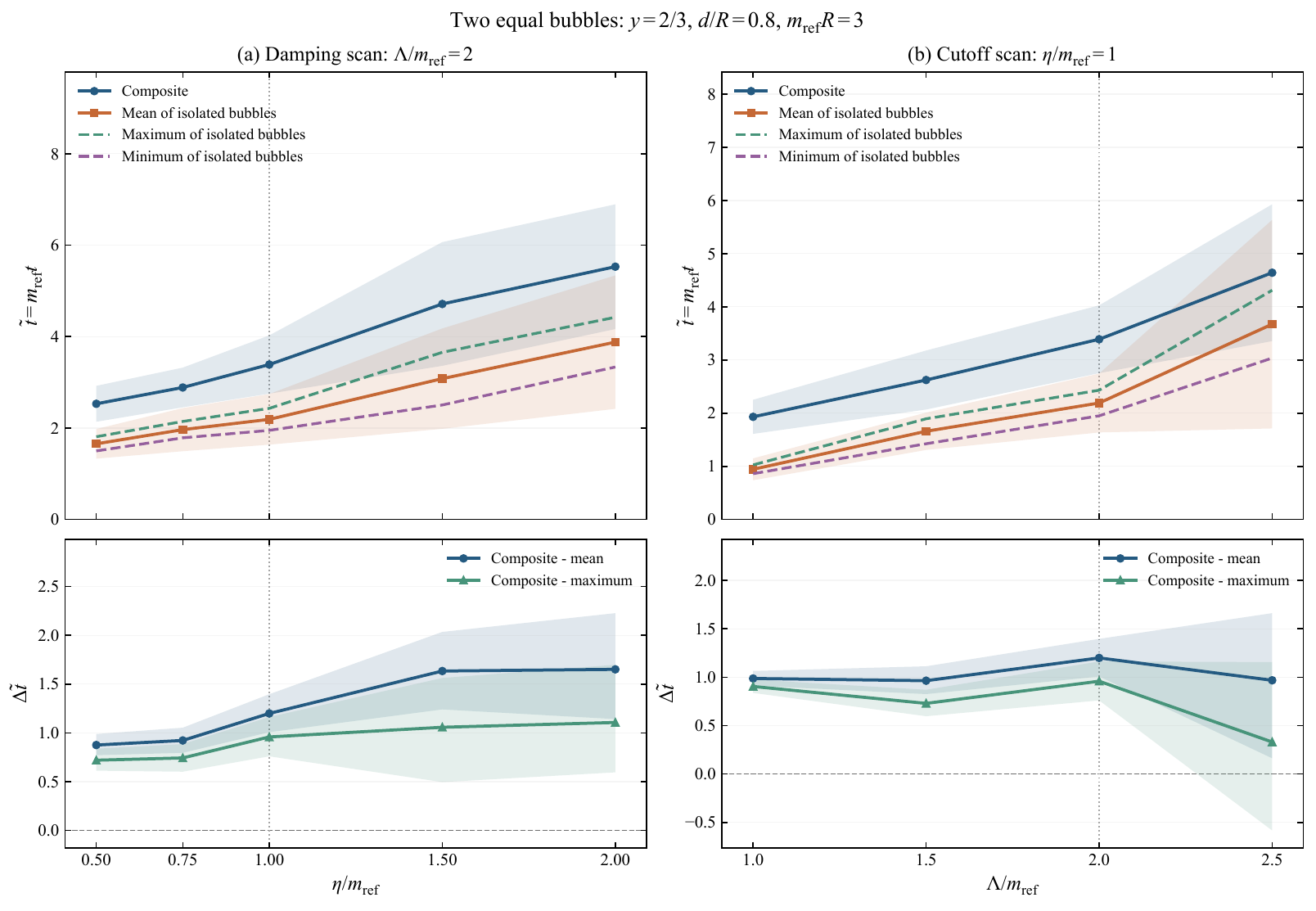}
\caption{
Sensitivity of two-bubble lifetimes to damping (left) and the Fourier
cutoff (right), at $y=2/3$, $d/R=0.8$, and $\tilde R_1=\tilde R_2=3$.
The unvaried parameter is fixed at $\tilde\Lambda=2$ or $\tilde\eta=1$,
respectively. Upper panels show ensemble mean lifetimes in
$\tilde t=\mref t$, with one-standard-deviation bands for the composite
and mean isolated references; the maximum and minimum isolated
references are averaged within the same ensembles.
Lower panels show paired lifetime differences with pointwise
$95\%$ bootstrap confidence intervals. Each point contains 30 experiments,
and vertical dotted lines mark the baseline parameters.
}
\label{fig:parameter_robustness}
\end{figure*}

To assess parameter robustness, we use the equal-width pair at
$y=2/3$ and $d/R=0.8$, varying
$\tilde\eta=0.5,0.75,1,1.5,2$ at fixed $\tilde\Lambda=2$, and
$\tilde\Lambda=1,1.5,2,2.5$ at fixed $\tilde\eta=1$.
All other input parameters and the tracking rule are held fixed.
Each point uses 30 matched noise seeds, with the same spatial noise
shared by the composite and isolated branches within each experiment.
The noise amplitude changes consistently with the damping, while a
cutoff change is applied to both the initial profiles and the subsequent
projected evolution at fixed potential coefficients.

Fig.~\ref{fig:parameter_robustness} shows that the absolute lifetimes
vary with both parameters, but the composite mean remains above the maximum
isolated lifetime throughout these scans. The pointwise $95\%$ confidence
intervals of the paired differences remain positive, supporting the
robustness of this qualitative lifetime ordering at the tested benchmark.

\begin{acknowledgments}
The numerical calculations in this work were carried out on the High-Performance Computing Platform at the Center for Theoretical Physics, Henan Normal University. This work was supported by the National Natural Science Foundation of China (Grant Nos.~12335005,12605177), the PI Research Fund of Henan Normal University (Grant No. 5101029470335), the National Natural Science Foundation of Henan province(Grant No. 262300421233 and 27AQ140008).
\end{acknowledgments}

\bibliographystyle{apsrev4-2}
\bibliography{references}
\end{document}